\documentclass[final,5p,times,twocolumn]{elsarticle}
\usepackage{lineno,hyperref,amsmath,amsthm,amssymb,amsfonts,ragged2e,color,subfig}
\journal{``Physics of Plasmas"}
\begin{document}
\begin{frontmatter}
\title{Modulational instability of ion-acoustic waves and associated  envelope solitons in a multi-component plasma}
\author{S. Banik$^{*,1,2}$, N.M. Heera$^{**,1}$, T. Yeashna$^{***,1}$, M.R. Hassan$^{\dag,1}$, R.K. Shikha$^{\ddag,1}$,\\
 N.A. Chowdhury$^{\S,3}$,  A. Mannan$^{\ddag\ddag,1,4}$, and A.A. Mamun$^{\S\S,1}$}
\address{$^{1}$Department of Physics, Jahangirnagar University, Savar, Dhaka-1342, Bangladesh\\
$^2$ Health Physics Division, Atomic Energy Centre, Dhaka-1000, Bangladesh\\
$^3$ Plasma Physics Division, Atomic Energy Centre, Dhaka-1000, Bangladesh\\
$^4$ Institut f\"{u}r Mathematik, Martin Luther Universit\"{a}t Halle-Wittenberg, Halle, Germany\\
e-mail: $^*$bsubrata.37@gmail.com, $^{**}$heera112phys@gmail.com, $^{***}$yeashna147phy@gmail.com,\\
$^{\dag}$hassan148phy@gmail.com, $^{\ddag}$shikha261phy@gmail.com, $^{\S}$nurealam1743phy@gmail.com,\\
$^{\ddag\ddag}$abdulmannan@juniv.edu, $^{\S\S}$mamun\_phys@juniv.edu}
\begin{abstract}
A generalized plasma model having warm ions, iso-thermal electrons, super-thermal electrons and
positrons is considered to theoretically investigate
the modulational instability (MI) of  ion-acoustic waves (IAWs).  A  standard
nonlinear Schr\"{o}dinger equation is derived by applying reductive
perturbation method to study the MI of IAWs. It is observed that the MI
criteria of the IAWs are significantly modified by various plasma parameters.
The present results should be useful in understanding the conditions for MI of IAWs which are relevant to both space
and laboratory plasma system.
\end{abstract}
\end{frontmatter}
\section{Introduction}
The co-existence of electrons and positrons in an electron-positron-ion
(EPI) plasma medium (EPIPM) which creates a new field of research for the
physicists to understand the nonlinear collective features of EPIPM by
considering waves dynamics, namely, ion-acoustic (IA)
waves (IAWs) \cite{Panwar2014,Kourakis2003,Alinejad2014,Rehman2016,
Shahmansouri2013,Shalini2015,Baluku2012}, positron-acoustic waves (PAWs),
electron-acoustic waves (EAWs) \cite{Baluku2011}, and IA rogue waves (IARWs)
as well as their associated nonlinear structures such as solitons \cite{Rehman2016,
Shahmansouri2013}, dark and bright envelope solitons \cite{Kourakis2003}, shocks,
rogue waves \cite{Shalini2015}, double layers has been identified by the
THEMIS mission \cite{Ergun1998} and Viking Satellite \cite{Temerin1982} in both
space (viz., Saturn's magnetosphere \cite{Panwar2014},  early universe \cite{Rehman2016},
solar atmosphere \cite{Panwar2014}, active galactic nuclei \cite{Chowdhury2017},
pulsar magnetosphere \cite{Rehman2016}, and polar regions of neutron
stars \cite{Chowdhury2018}, etc.) and laboratory environments (viz., high intensity
laser irradiation \cite{Rehman2016}, semiconductor plasmas \cite{Chowdhury2017},
hot cathode discharge \cite{Rehman2016}, and magnetic confinement systems
\cite{Chowdhury2017}, etc.).

Two temperature electrons (hot and cold) have been identified by the Voyager
PLS \cite{Sittler1983} and Cassini CAPS \cite{Young2005} observations in
Saturn's magnetosphere, and successively  have been verified by several satellite
missions, viz., Viking Satellite \cite{Temerin1982}, FAST Auroral Snapshot (FAST)
at the auroral region \cite{Pottelette1999} and THEMIS mission \cite{Ergun1998},
and are governed by the super-thermal kappa/$\kappa$-distribution rather than
well-known Maxwellian distribution, and have also been considered by  many
authors for analysing the propagation of the nonlinear electrostatic waves
\cite{Panwar2014,Alinejad2014,Baluku2011,Vasyliunas1968}. The super-thermal parameter
($\kappa$) in $\kappa$-distribution represents super-thermality of plasma species, and
the small values of $\kappa$ determine the large deviation of the plasma species from
thermally equilibrium state of plasma system while for the large values of $\kappa$,
the plasma system coincides with the Maxwellian distribution. Shahmansouri and Alinejad
\cite{Shahmansouri2013} considered three components plasma model having two temperature
super-thermal electrons and cold ions, and investigated IA solitary waves, and confirmed
that the existence of both compressive and rarefactive solitary structures in presence of
the two temperature super-thermal electrons. Baluku and Helberg theoretically and
numerically analyzed IA solitons in presence of two temperature super-thermal electrons.
Panwar \textit{et al.} \cite{Panwar2014} have demonstrated IA cnoidal waves in a three components
plasma medium having inertial cold ion and inertialess two temperature $\kappa$-distributed electrons,
and found that cold electron's super-thermality increases the height of the cnoidal wave.

The  MI of wave packets has been considered the basic platform for the
formation of bright and dark envelope solitons in plasmas, and has also been mesmerized a
number of authors to investigate the MI as well as bright and dark envelope solitons
in inter-disciplinary field of nonlinear-sciences, viz., oceanic wave \cite{Shalini2015}, fibre
telecommunications \cite{Kourakis2003}, optics \cite{Shalini2015}, and space plasma
\cite{Alinejad2014}, etc. The intricate mechanism of the MI of various waves (viz., IAWs,
EAWs, and PAWs \cite{C4}, etc.) and the formation of the electrostatic envelope solitonic
solitons has been governed by the standard nonlinear Schr\"{o}dinger equation (NLSE) \cite{C5,C6,C7,C8,Ahmed2018}.
Kourakis and Shukla \cite{Kourakis2003} investigated the MI of the
IAWs in a super-thermal plasma having inertial cold ion and inertialess cold and hot
electrons. Alinejad \textit{et al.} \cite{Alinejad2014} studied the stability conditons
of the IAWs in presence of the super-thermal electrons, and found that the stable region of the IAWs
decreases with the number density of the cold electrons. Ahmed  \textit{et al.} \cite{Ahmed2018}
studied the stability of IAWs, and observed that the critical
wave number $k_c$ decreases with the increase in the value of $\kappa$.

The manuscript is organized in the following order: The governing equations
of plasma model are presented in Sec. \ref{2sec:Governing Equations}. The
derivation of NLSE by using reductive perturbation method (RPM) is represented
in Sec. \ref{2sec:Derivation of NLSE}. The MI of IAWs is given in Sec. \ref{2sec:Modulational instability}.
Envelope solitons are provided in Sec. \ref{2sec:Envelope Soliton}. The results
and discussion is given in Sec. \ref{2sec:Results and discussion}. Finally, the conclusion is presented
in  Sec. \ref{2sec:Conclusion}.
\section{Governing Equations}
\label{2sec:Governing Equations}
We consider a four component unmagnetized plasma model consisting of
warm ions (with charge $q_+=Z_+e$; mass $m_+$), $\kappa$-distributed super-thermal
electrons (with charge $q_{e1}=-e$; mass $m_{e1}$), iso-thermal electrons
(with charge $q_{e2}=-e$; mass $m_{e2}$), and super-thermal $\kappa$-distributed positrons (with charge $q_{p}=e$; mass $m_{p}$).
The overall charge neutrality condition for this plasma model can be written as
$Z_+ n_{+0} + n_{p0} = n_{e10} + n_{e20}$; where $n_{+0}$, $n_{p0}$, $n_{e10}$, and $n_{e20}$ are the equilibrium
number densities of warm ions, super-thermal positrons and electron, and iso-thermal electrons.
Now, the basic set of normalized equations can be written in the following form
\begin{eqnarray}
&&\hspace*{-1.3cm}\frac{\partial n_+}{\partial t}+\frac{\partial}{\partial x}(n_+ u_+)=0,
\label{2eq:1}\\
&&\hspace*{-1.3cm}\frac{\partial u_+}{\partial t} + u_+\frac{\partial u_+}{\partial x}+ \alpha n_+\frac{\partial n_+}{\partial x}=-\frac{\partial \phi}{\partial x},
\label{2eq:2}\\
&&\hspace*{-1.3cm}\frac{\partial^2 \phi}{\partial x^2}+ n_+ = \mu_1 n_{e1}+\mu_2 n_{e2}-(\mu_1+\mu_2-1)n_P,
\label{2eq:3}\
\end{eqnarray}
where $n_+$ is the number density of inertial warm ions normalized by its equilibrium value $n_{+0}$;
$u_i$ is the ion fluid speed normalized by the IAW speed $C_+=\sqrt{Z_+k_BT_{e1}/m_+}$ (with $T_{e1}$ being
the $\kappa$-distributed electron temperature, $m_i$ being the ion rest mass, and $k_B$ being the
Boltzmann constant); $\phi$ is the electrostatic wave potential normalized by $k_B T_{e1}/e$ (with $e$ being
the magnitude of single electron charge); the time and space variables are normalized by
$\omega_{p+}=\sqrt{4\pi e^2Z_+^2n_{+0}/m_+}$ and $\lambda_{D+}=\sqrt{k_B T_{e1}/4\pi e^2Z_+n_{+0}}$, respectively. The pressure term of the
ion can be written as $P_+=P_{+0}(N_+/n_{+0})^\gamma$ with $P_{+0}=n_{+0}k_BT_+$ being the equilibrium pressure of
the ion, and $T_+$ being the temperature of warm ion, and $\gamma=(N + 2)/N$ (where $N$ is the degrees
of freedom and for one-dimensional case $N=1$, and then $\gamma=3$). Other parameters are defined as
$\alpha=3T_+/Z_+T_{e1}$, $\mu_1=n_{e10}/Z_+n_{+0}$, $\mu_2=n_{e20}/Z_+n_{+0}$, and $\mu_p=n_{p0}/Z_+n_{+0}$.

The expression for number density of super-thermal electron following the
$\kappa$-distribution \cite{Alinejad2014} can be expressed as
\begin{eqnarray}
&&\hspace*{-1.3cm}n_{e1}=\left[1-\frac{\phi}{(\kappa-3/2)}\right]^{(-\kappa+1/2)}
\nonumber\\
&&\hspace*{-0.5cm}= 1+ F_1 \phi+ F_2 \phi^2+ F_3 \phi^3+ \cdot\cdot\cdot,
\label{2eq:4}
\end{eqnarray}
where
\begin{eqnarray}
&&\hspace*{-1.3cm}F_1=(2\kappa-1)/(2\kappa-3),
\nonumber\\
&&\hspace*{-1.3cm}F_2=[(2\kappa-1)(2\kappa+1)]/2(2\kappa-3)^2,
\nonumber\\
&&\hspace*{-1.3cm}F_3=[(2\kappa-1)(2\kappa+1)(2\kappa+3)]/6(2\kappa-3)^3.
\nonumber\
\end{eqnarray}
The expression for number density of super-thermal positron following the
$\kappa$-distribution \cite{Alinejad2014} can be expressed as
\begin{eqnarray}
&&\hspace*{-1.3cm}n_p=\Big[1+\frac{\delta\phi}{(\kappa-3/2)}\Big]^{(-\kappa+1/2)}
\nonumber\\
&&\hspace*{-0.5cm}= 1- F_4\phi+ F_5\phi^2- F_6 \phi^3+\cdot\cdot\cdot,
\label{2eq:5}
\end{eqnarray}
where $F_4=F_1\delta$, $F_5=F_2\delta^2$, $F_6=F_3\delta^3$, and $\delta=T_{e1}/{T_p}$
(with $T_p$ is the super-thermal positron temperature).
The expression for the number density of iso-thermal electron can be expressed as
\begin{eqnarray}
&&\hspace*{-1.3cm}n_{e2}=\exp(\lambda\phi)= 1+ \lambda\phi+ \frac{\lambda^2}{2} \phi^2+ \frac{\lambda^3}{6} \phi^3+ \cdot\cdot\cdot,
\label{2eq:6}
\end{eqnarray}
where $\lambda=T_{e1}/T_{e2}$ (with $T_{e1}>T_{e2}$). Now, by substituting Eqs. \eqref{2eq:4}-
\eqref{2eq:6} into Eq. \eqref{2eq:3}, and expanding up to third order in $\phi$, we get
\begin{eqnarray}
&&\hspace*{-1.3cm}\frac{\partial^2\phi}{\partial x^2} + n_+ = 1+F_7 \phi + F_8 \phi^2 + F_9 \phi^3+ \cdot\cdot\cdot,
\label{2eq:7}
\end{eqnarray}
where
\begin{eqnarray}
&&\hspace*{-1.3cm}F_7=\mu_1 F_1+ \mu_2 \lambda + (\mu_1+ \mu_2 -1) F_4,
\nonumber\\
&&\hspace*{-1.3cm}F_8=\mu_1 F_2+ \mu_2 \lambda^2/2 + (\mu_1+ \mu_2 -1) F_5,
\nonumber\\
&&\hspace*{-1.3cm}F_9=\mu_1 F_3+ \mu_2\lambda^3/6 + (\mu_1+ \mu_2 -1) F_6.
\nonumber\
\end{eqnarray}
The terms containing $F7$, $F8$, and $F9$ in the right-hand side
of Eq. \eqref{2eq:7} are the contribution of inertialess electrons and positrons.
\section{Derivation of the NLSE}
\label{2sec:Derivation of NLSE}
We can employ the RPM to derive the NLSE and hence to study the MI of IAWs.
So, the stretched co-ordinates can be written as
\begin{eqnarray}
&&\hspace*{-1.3cm}\xi=\epsilon(x-v_gt),
\label{2eq:8}\\
&&\hspace*{-1.3cm}\tau=\epsilon^2 t,
\label{2eq:9}
\end{eqnarray}
where $v_g$ and $\epsilon$ are denoted as the group speed and smallness parameter, respectively.
So, the dependent variables  can be written as
\begin{eqnarray}
&&\hspace*{-1.3cm}\Pi(x,t)=\Pi_0+\sum_{m=l}^{\infty}\epsilon^{(m)} \sum_{l=-\infty}^{\infty} \Pi_{l}^{(m)} (\xi,\tau)
~\mbox{exp}[i l(kx-\omega t)],
\label{2eq:10}
\end{eqnarray}
where $\Pi_l^{(m)}$= $[n_{+l}^{(m)}$, $u_{+l}^{(m)}$, $\phi_l^m]$, $\Pi_0$ = $[1,0,0]^T$,
and $\omega$ is the angular frequency. The carrier wave number is represented by $k$.
The derivative operators can be written as
\begin{eqnarray}
&&\hspace*{-1.3cm}\frac{\partial}{\partial t}\rightarrow\frac{\partial}{\partial t}-\epsilon v_g \frac{\partial}{\partial\xi}+\epsilon^2\frac{\partial}{\partial\tau},
\label{2eq:11}\\
&&\hspace*{-1.3cm}\frac{\partial}{\partial x}\rightarrow\frac{\partial}{\partial x}+\epsilon\frac{\partial}{\partial\xi}.
\label{2eq:12}
\end{eqnarray}
Now, by substituting Eqs. \eqref{2eq:8}-\eqref{2eq:12}, into Eqs. \eqref{2eq:1}, \eqref{2eq:2},
and \eqref{2eq:7}, and selecting the terms containing $\epsilon$, the first order ($m=1$
and $l=1$) reduced equations can be written as
\begin{eqnarray}
&&\hspace*{-1.3cm}iku_{+1}^{(1)}-i\omega n_{+1}^{(1)}=0,
\label{2eq:13}\\
&&\hspace*{-1.3cm}ik\phi_1^{(1)}+ik\alpha n_{+1}^{(1)}-i\omega u_{+1}^{(1)}=0,
\label{2eq:14}\\
&&\hspace*{-1.3cm}n_{+1}^{(1)}-k^2 \phi_1^{(1)}-F_7 \phi_1^{(1)}=0,
\label{2eq:15}
\end{eqnarray}
these equations provide the dispersion relation for IAWs
\begin{eqnarray}
&&\hspace*{-1.3cm}\omega^2=\frac{k^2}{k^2+F_7}+k^2 \alpha,
\label{2eq:16}
\end{eqnarray}
For second order harmonics, equations can be found from the next order of $\epsilon$ (with $m=2$ and $l=1$) as
\begin{eqnarray}
&&\hspace*{-1.3cm}n_{+1}^{(2)}=\frac{k^2}{(\omega^2-\alpha k^2)} \phi_1^{(2)}+\frac{2ik\omega (v_g k-\omega)}{(\omega^2-k^2 \alpha)^2} \frac{\partial \phi_1^{(1)}}{\partial \xi},
\label{2eq:17}\\
&&\hspace*{-1.3cm}u_{+1}^{(2)}=\frac{k\omega}{(\omega^2- \alpha k^2)} \phi_1^{(2)}+ i\frac{(v_g k-\omega)(\omega^2+\alpha k^2)}{(\omega^2-k^2 \alpha)^2} \frac{\partial \phi_1^{(1)}}{\partial \xi},
\label{2eq:18}
\end{eqnarray}
with the compatibility condition, we have obtained the group speed of IAWs as
\begin{eqnarray}
&&\hspace*{-1.3cm}v_g=\frac{\partial\omega}{\partial k}=\frac{\omega^2-(\omega^2-k^2 \alpha)^2}{k \omega},
\label{2eq:19}
\end{eqnarray}
when $m=2$ with $l=2$, second order harmonic amplitudes are found for the coefficient of $\epsilon$ in terms of $\vert \phi_1^{(1)} \vert^2$ as
\begin{eqnarray}
&&\hspace*{-1.3cm}n_{+2}^{(2)}=F_{10}\vert \phi_1^{(1)}\vert^2,
\label{2eq:20}\\
&&\hspace*{-1.3cm}u_{+2}^{(2)}=F_{11}\vert \phi_1^{(1)}\vert^2,
\label{2eq:21}\\
&&\hspace*{-1.3cm}\phi_{2}^{(2)}=F_{12}\vert \phi_1^{(1)}\vert^2,
\label{2eq:22}
\end{eqnarray}
where
\begin{eqnarray}
&&\hspace*{-1.3cm}F_{10}= \frac{k^2\{\alpha k^4+3\omega^2 k^2+ 2 F_{12}(\omega^2-k^2 \alpha)^2\}}{2(\omega^2-k^2 \alpha)^3},
\nonumber\\
&&\hspace*{-1.3cm}F_{11}=\frac{F_{10}(\omega^5-2\omega^3 k^2\alpha+\omega k^4 \alpha^2)-\omega k^4}{k(\omega^2-k^2 \alpha)^2},
\nonumber\\
&&\hspace*{-1.3cm}F_{12}=\frac{k^4(3\omega^2+\alpha k^2)-2 F_8 (\omega^2-k^2\alpha)^3}{2(\omega^2-k^2\alpha)^2[4\omega^2 k^2-4\alpha k^4+F_7 \omega^2-\alpha F_7 k^2-k^2]}.
\nonumber
\end{eqnarray}
Now, we consider the third-order expression for $m=3$ with $l=0$ and $m=2$ with $l=0$ that leads to zeroth harmonic modes. Thus, we obtain
\begin{eqnarray}
&&\hspace*{-1.3cm}n_{+0}^{(2)}=F_{13}\vert \phi_1^{(1)}\vert^2,
\label{2eq:23}\\
&&\hspace*{-1.3cm}u_{+0}^{(2)}=F_{14}\vert \phi_1^{(1)}\vert^2,
\label{2eq:24}\\
&&\hspace*{-1.3cm}\phi_{0}^{(2)}=F_{15}\vert \phi_1^{(1)}\vert^2,
\label{2eq:25}
\end{eqnarray}
where
\begin{eqnarray}
&&\hspace*{-1.3cm}F_{13}=\frac{k^2(2\omega v_g k+\alpha k^2+\omega^2)+F_{15}(\omega^4-2\omega^2 k^2\alpha+k^4\alpha^2)}{(v_g^2-\alpha)(\omega^2-k^2\alpha)^2},
\nonumber\\
&&\hspace*{-1.3cm}F_{14}=\frac{F_{13}(v_g\omega^4-2\omega^2 k^2 v_g \alpha+v_g k^4\alpha^2)-2\omega k^3}{(\omega^2-k^2\alpha)^2},
\nonumber\\
&&\hspace*{-1.3cm}F_{15}=\frac{k^2(2\omega v_g k+\alpha k^2+\omega^2)-2 F_8 (\omega^2-k^2\alpha)^2 (v_g^2-\alpha)}{(\omega^2-k^2\alpha)^2 [F_7(v_g^2-\alpha)-1]}.
\nonumber
\end{eqnarray}
Now, we develop the standard NLSE by substituting all the above equations into third order harmonic modes ($m=3$ with $l=1$) as
\begin{eqnarray}
&&\hspace*{-1.3cm}i\frac{\partial\Phi}{\partial \tau}+P\frac{\partial^2 \Phi}{\partial \xi^2}+Q|\Phi|^2\Phi=0,
\label{2eq:26}
\end{eqnarray}
where $\Phi$=$\phi_1^{(1)}$ for simplicity. In Eq. \eqref{2eq:26}, $P$ is the dispersion coefficient which can be written as
\begin{eqnarray}
&&\hspace*{-1.3cm}P=\frac{(\omega^2-k^2\alpha)(4\alpha k^2\omega-3v_gk\omega^2-v_g k^3\alpha)}{2\omega^2 k^2},
\label{2eq:27}
\end{eqnarray}
and $Q$ is the nonlinear coefficient which can be written as
\begin{eqnarray}
&&\hspace*{-1.3cm}Q=\frac{1}{2\omega k^2}\times[(\omega^2-k^2\alpha)^2\{3 F_9+2 F_8 (F_{12}+F_{15})\}
\nonumber\\
&&\hspace*{-0.5cm}-k^2\{(\alpha k^2+\omega^2)(F_{10}+F_{13})-2\omega k(F_{11} +F_{14})\}].
\label{2eq:28}
\end{eqnarray}
The space and time evolution of the IAWs in plasma medium are directly governed by the  dispersion
($P$) and nonlinear  ($Q$) coefficients of NLSE, and are indirectly governed by different plasma parameters such as
as $\alpha$, $\delta$,  $\lambda$, $\kappa$, $\mu_1$, and $\mu_2$. Thus, these plasma parameters significantly
affect the stability conditions of IAWs.
\begin{figure}
\centering
\includegraphics[width=70mm]{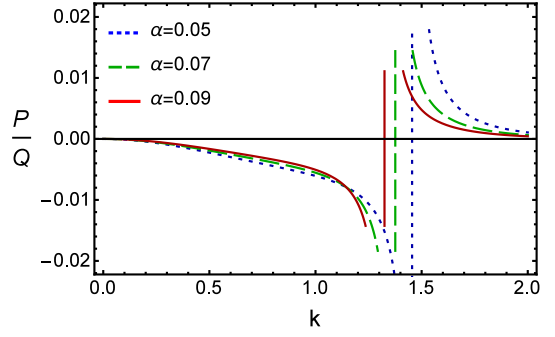}
\caption{The variation of $P/Q$ with $k$ for different
values of $\alpha$ when $\delta=1.2$, $\lambda=1.5$,
$\kappa=2$, $\mu_1=0.7$, and $\mu_2=0.5 $.}
\label{2Fig:F1}
\end{figure}
\begin{figure}
\centering
\includegraphics[width=70mm]{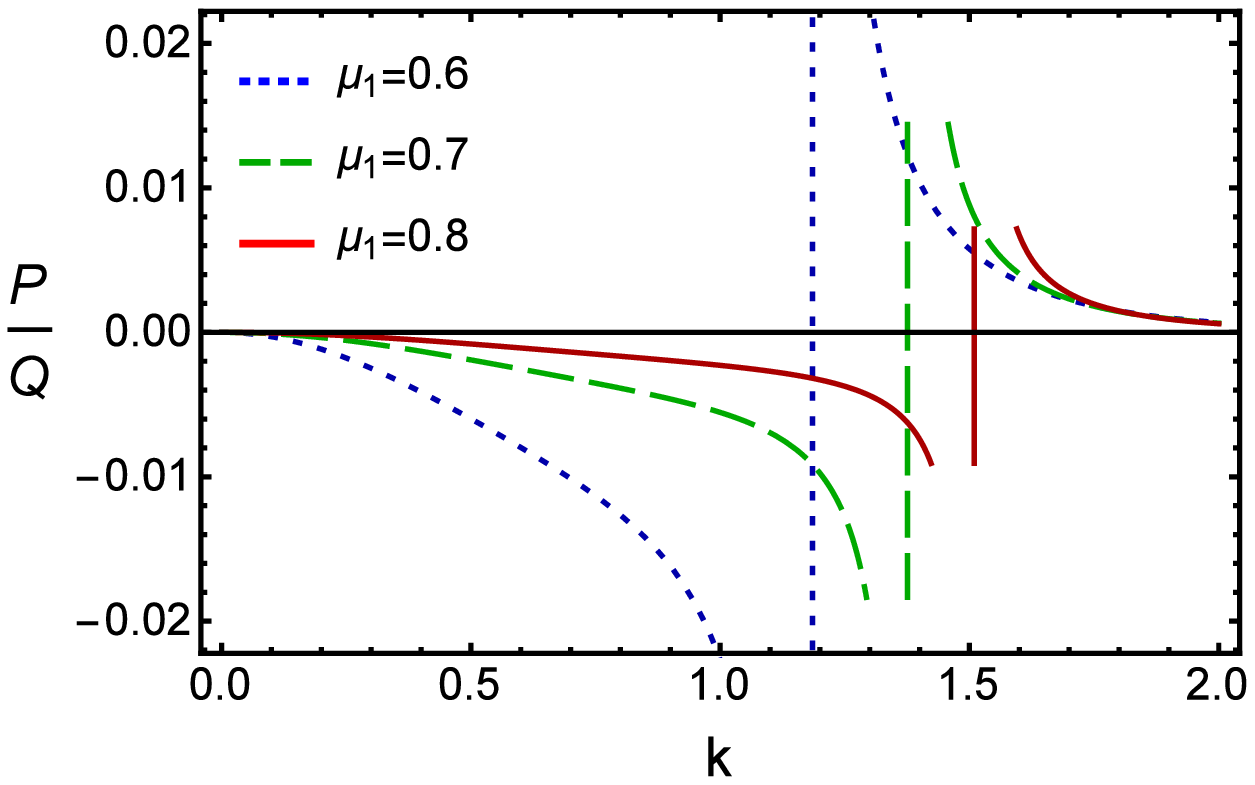}
\caption{The variation of $P/Q$ with $k$ for different
values of $\mu_1$ when $\alpha=0.07$, $\delta=1.2$,
$\lambda=1.5$, $\kappa=2$, and $\mu_2=0.5$.}
\label{2Fig:F2}
\end{figure}
\begin{figure}
\centering
\includegraphics[width=70mm]{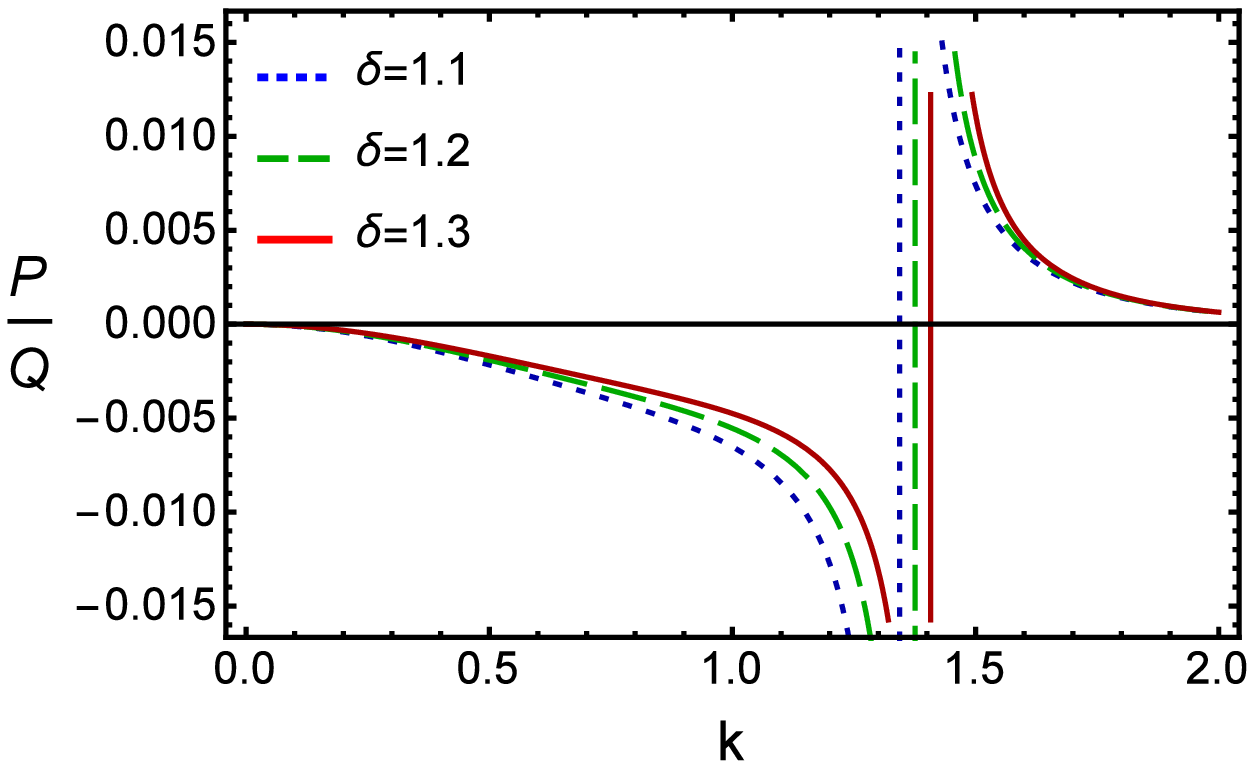}
\caption{The variation of $P/Q$ with $k$ for different
values of $\delta$ when $\alpha=0.07$, $\lambda=1.5$,
$\kappa=2$, $\mu_1=0.7$, and $\mu_2=0.5$.}
\label{2Fig:F3}
\end{figure}
\begin{figure}
\centering
\includegraphics[width=70mm]{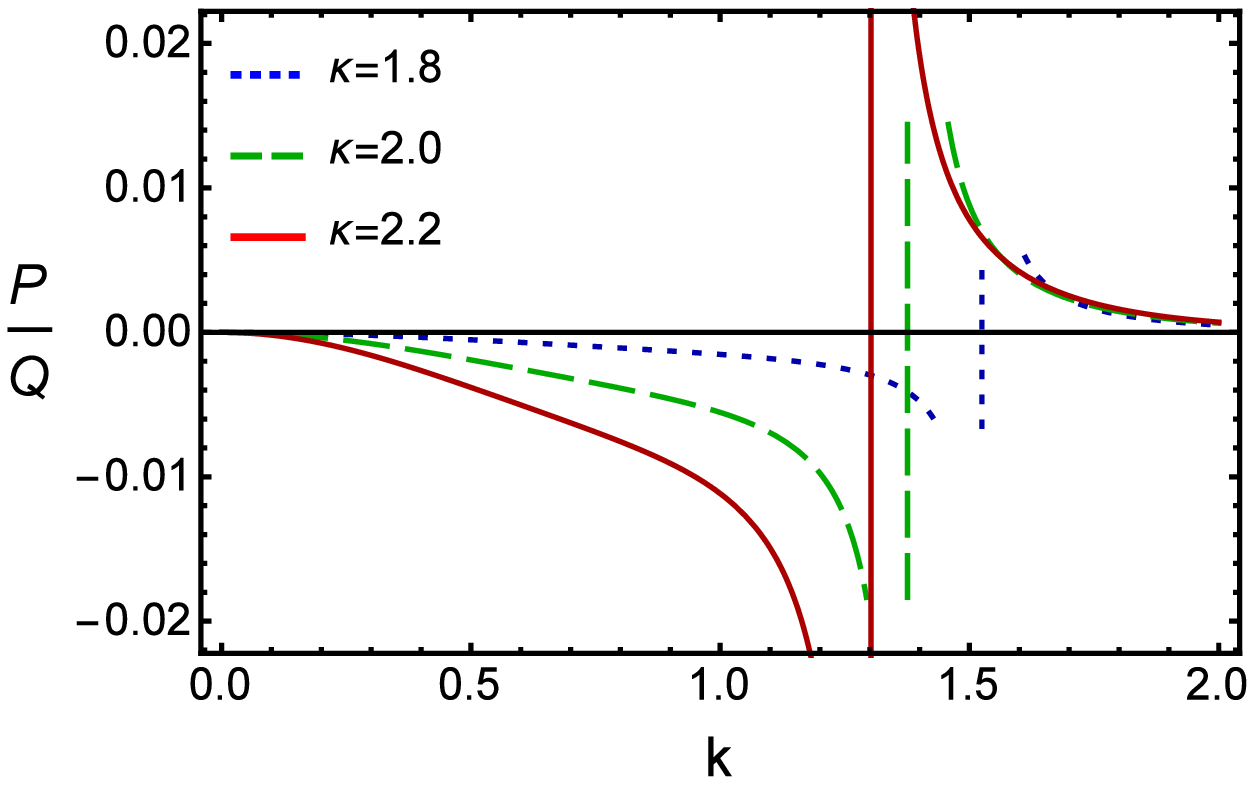}
\caption{The variation of $P/Q$ with $k$ for different
values of $\kappa$ when $\alpha=0.07$, $\delta=1.2$
$\lambda=1.5$, $\mu_1=0.7$, and $\mu_2=0.5$.}
\label{2Fig:F4}
\end{figure}
\begin{figure}
\centering
\includegraphics[width=70mm]{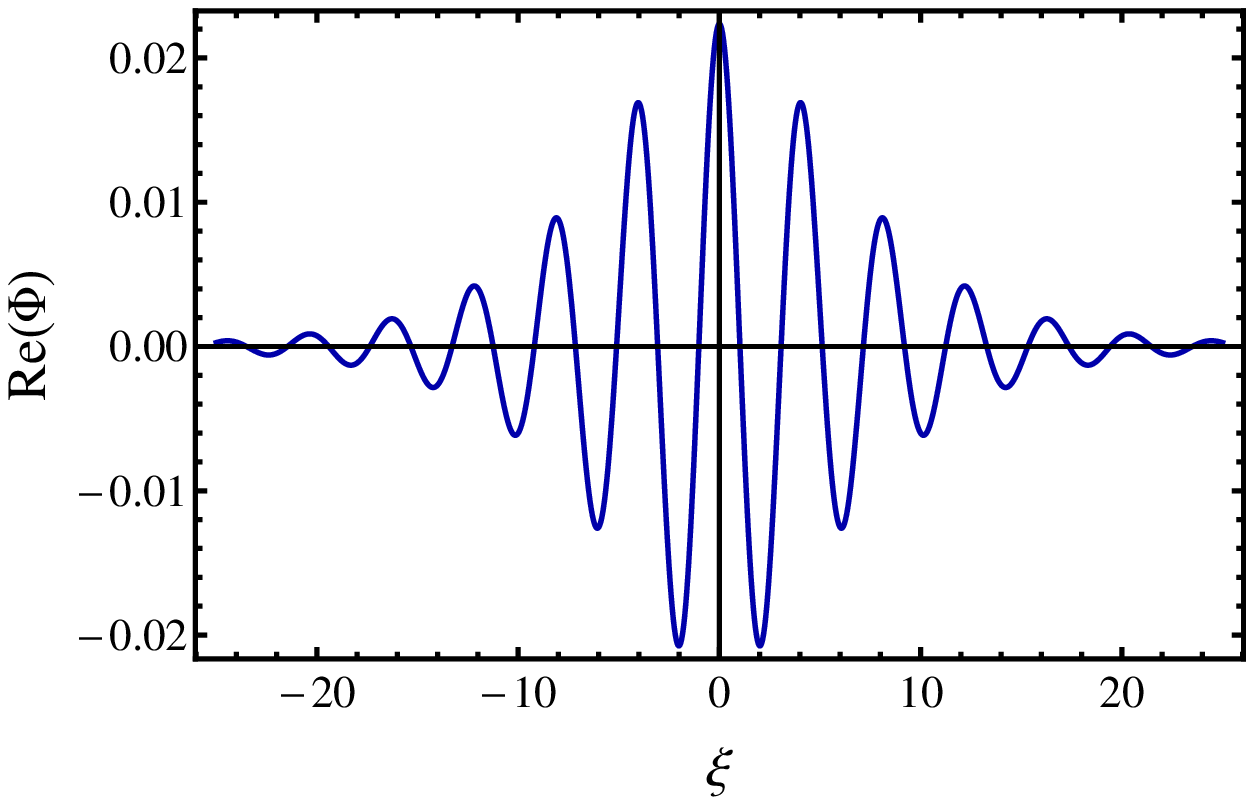}
\caption{The bright envelope solitons for $\alpha=0.07$, $\delta=1.2$, $\lambda=1.5$, $\kappa=2$,
$\mu_1=0.8$, $\mu_2=0.5$, $\tau=0$, $\psi_0=0.0005$, $U=0.2$, $\Omega_0=0.4$, and $k=1.6$.}
\label{2Fig:F5}
\end{figure}
\begin{figure}
\centering
\includegraphics[width=70mm]{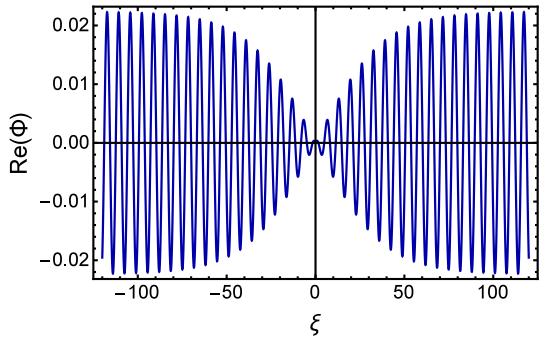}
\caption{The dark envelope solitons for  $\alpha=0.07$, $\delta=1.2$, $\lambda=1.5$,
$\kappa=2$, $\mu_1=0.8$, $\mu_2=0.3$, $\tau=0$, $\psi_{0}=0.0005$, $U=0.2$, $\Omega_0=0.4$, and $k=1.2$.}
\label{2Fig:F6}
\end{figure}
\section{Modulational instability}
\label{2sec:Modulational instability}
The stable and unstable parametric regimes of IAWs are
organised by the sign of $P$ and $Q$ of Eq. \eqref{2eq:26} \cite{Kourakis2005,Sultana2011,C9,C10,C11,C12,C13,C14}.
When $P$ and $Q$ have the same sign (i.e., $P/Q > 0$),
the evolution of IAWs amplitude is modulationally
unstable in the presence of external perturbations.
On the other hand, when $P$ and $Q$ have opposite signs
(i.e., $P/Q < 0$), the IAWs are modulationally stable in
the presence of external perturbations. The plot of $P/Q$
against $k$ yields stable and unstable parametric regimes
of the IAWs. The point, at which the transition of $P/Q$
curve intersects with the $k$-axis, is known as the threshold
or critical wave number $k~(= k_c)$ \cite{Kourakis2005,Sultana2011,C9,C10,C11,C12,C13,C14}.
\section{Envelope Solitons}
\label{2sec:Envelope Soliton}
The bright (when $P/Q>0$) and dark (when $P/Q<0$) envelope
solitonic solutions, respectively, can be written as \cite{Kourakis2005,Sultana2011}
\begin{eqnarray}
&&\hspace*{-1.3cm}\Phi(\xi,\tau)=\left[\psi_0~\mbox{sech}^2 \left(\frac{\xi-U\tau}{W}\right)\right]^\frac{1}{2}
\nonumber\\
&&\hspace*{-0.01cm}\times \exp \left[\frac{i}{2P}\left\{U\xi+\left(\Omega_0-\frac{U^2}{2}\right)\tau \right\}\right],
\label{1eq:29}\\
&&\hspace*{-1.3cm}\Phi(\xi,\tau)=\left[\psi_{0}~\mbox{tanh}^2 \left(\frac{\xi-U\tau}{W}\right)\right]^\frac{1}{2}
\nonumber\\
&&\hspace*{-0.01cm}\times \exp \left[\frac{i}{2P}\left\{U\xi-\left(\frac{U^2}{2}-2 P Q \psi_{0}\right)\tau \right\}\right],
\label{1eq:30}
\end{eqnarray}
where $\psi_0$ is the amplitude of localized pulse for both bright and dark
envelope soliton, $U$ is the propagation speed of the localized pulse, $W$ is the soliton width, and
$\Omega_0$ is the oscillating frequency at $U=0$. The soliton width $W$ and the maximum amplitude  $\psi_0$
are related by $W=\sqrt{2\mid P/Q\mid/\psi_0}$. We have observed the bright and dark envelope solitons in Fig. \ref{2Fig:F5}-\ref{2Fig:F6}.
\section{Results and discussion}
\label{2sec:Results and discussion}
We have graphically examined the effects of the temperature of  warm ion and
super-thermal electron as well as the charge state
of the warm ion in recognizing the stable and unstable domains
of the IAWs in Fig. \ref{2Fig:F1}, and it is clear from this figure
that (a) the stable domain decreases with the increase in the
value of warm ion temperature but increases with the increase
of the value of super-thermal electron temperature when the
charge state of the warm ion remains constant; (b) the stable
domain increases with $Z_+$ for constant value of $T_+$ and $T_{e1}$ (via $\alpha=3T_+/Z_+T_{e1}$).
So, the charge state and temperature of warm ion
play an opposite role in manifesting the stable and unstable domains of IAWs.

Both stable (i.e., $k<k_c$) and unstable (i.e., $k>k_c$) domains for the IAWs can
be observed from Fig. \ref{2Fig:F2}, and it is obvious from this figure that
(a) when $\mu_1=0.6$, $0.7$, and $0.8$, then the
corresponding value of $k_c$ is $1.20$ (dotted blue curve), $1.40$
(dashed green curve), and $1.50$ (solid red curve); (b) $k_c$ is shifted to higher values with the
increase (decrease) of $n_{e10}$ ($n_{+0}$) when the value of $Z_+$ is constant.
Finally, $\mu_1$ would cause to increase the stable domain of IAWs;

We have numerically analyzed the effect of temperature of the super-thermal electron and
super-thermal positron on the stability conditions of IAWs in Fig. \ref{2Fig:F3}.
It can be seen from this figure that the stable domain increases with the increase (decrease)
in the value of super-thermal electron (positron) temperature. Figure \ref{2Fig:F4} describes
the effects of super-thermality of plasma species in the stable and unstable parametric domains.
It is clear from this figure that for large values of $\kappa$, the IAWs become unstable for
small values of $k$  while for small values of $\kappa$, the IAWs become unstable for
large values of $k$.
\section{Conclusion}
\label{2sec:Conclusion}
We have studied an unmagnetized realistic space
plasma system consists of warm ions,  iso-thermal electrons, $\kappa$-distributed
electrons and positrons.  The RPM is used to derive the NLSE.
The existence of both stable and unstable regions of IAWs has been found,
and the interaction between the $k_c$ with various plasma parameters (i.e., $\alpha$, $\mu_1$,
$\delta$, and $\kappa$, etc.) has also been observed.
Finally, these results may be applicable
in understanding the conditions of the MI of IAWs
and associated envelope solitons in astrophysical environments.

\end{document}